\documentclass[conference,10pt]{IEEEtran}
\IEEEoverridecommandlockouts

\usepackage{epsfig,rotating,setspace,latexsym,amsmath,epsf,amssymb,amsfonts,bm,theorem,cite,algorithm,graphicx,epsf,authblk,epstopdf,color,algpseudocode,bbm}

\newtheorem{corollary}{Corollary}
\newtheorem{definition}{Definition}
\newtheorem{theorem}{Theorem}
\newtheorem{lemma}{Lemma}
\newenvironment{Proof}[1]{\medskip\par\noindent{\bf Proof:\,}\,#1}{{\mbox{\,$\blacksquare$}\par}}

\allowdisplaybreaks

\begin{document}

\title{Age-Minimal Transmission in Energy Harvesting Two-hop Networks\thanks{This work was supported by NSF Grants CNS 13-14733, CCF 14-22111, CCF 14-22129, and CNS 15-26608.}}

\author{Ahmed Arafa \quad Sennur Ulukus\\
\normalsize Department of Electrical and Computer Engineering\\
\normalsize University of Maryland, College Park, MD 20742\\
\normalsize \emph{arafa@umd.edu} \quad \emph{ulukus@umd.edu}}

\maketitle

\begin{abstract}
We consider an energy harvesting two-hop network where a source is communicating to a destination through a relay. During a given communication session time, the source collects measurement updates from a physical phenomenon and sends them to the relay, which then forwards them to the destination. The objective is to send these updates to the destination as {\it timely} as possible; namely, such that the total {\it age of information} is minimized by the end of the communication session, subject to energy causality constraints at the source and the relay, and data causality constraints at the relay.  Both the source and the relay use fixed, yet possibly different, transmission rates. Hence, each update packet incurs fixed non-zero transmission delays. We first solve the single-hop version of this problem, and then show that the two-hop problem is solved by treating the source and relay nodes as one combined node, with some parameter transformations, and solving a single-hop problem between that combined node and the destination.
\end{abstract}

\section{Introduction}

\begin{figure}[t]
\center
\includegraphics[scale=.8]{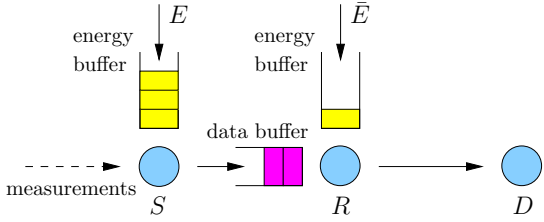}
\caption{Energy harvesting two-hop network. The source collects measurements and sends them to the destination through the relay.}
\label{fig_2hop_sys}
\vspace{-0.2in}
\end{figure}

A source node is collecting measurements from a physical phenomenon and sends updates to a destination through the help of a relay, see Fig.~\ref{fig_2hop_sys}. Both the source and the relay depend on energy harvested from nature to communicate. Updates need to be sent in a {\it timely} manner; namely, such that the total {\it age of information} is minimized by a given deadline. The age of information is defined as the time elapsed since the freshest update has reached the destination.

Power scheduling in energy harvesting communication systems has been extensively studied in the recent literature. Earlier works \cite{jingP2P, kayaEmax, omurFade, ruiZhangEH} consider the single-user setting under different battery capacity assumptions, with and without fading. References \cite{jingBC, omurBC, elifBC, jingMAC, kaya-interference} extend this to multiuser settings: broadcast, multiple access, and interference channels; and \cite{ruiZhangRelay, gunduz2hop, berkDiamond-jour, varan_twc_jour, arafa_baknina_twc_dec_proc} consider two-hop, relay, and two-way channels.

Minimizing the age of information metric has been studied mostly in a queuing-theoretic framework; \cite{yates_age_1} studies a source-destination link under random and deterministic service times. This is extended to multiple sources in \cite{yates_age_mac}. References \cite{ephremides_age_random, ephremides_age_management, ephremides_age_non_linear} consider variations of the single source system, such as randomly arriving updates, update management and control, and nonlinear age metrics. \cite{shroff_age_mdp} introduces penalty functions to assess age dissatisfaction; and \cite{shroff_age_multi_hop} 
shows that last-come-first-serve policies are optimal in multi-hop networks.

Our work is most closely related to \cite{yates_age_eh, elif_age_eh}, where age minimization in single-user energy harvesting systems is considered; the difference of these works from energy harvesting literature in \cite{jingP2P, kayaEmax, omurFade, ruiZhangEH, jingBC, omurBC, elifBC, jingMAC, kaya-interference, ruiZhangRelay, gunduz2hop, berkDiamond-jour, varan_twc_jour, arafa_baknina_twc_dec_proc} is that the objective is age of information as opposed to throughput or transmission completion time, and the difference of them from age minimization literature in \cite{yates_age_1, yates_age_mac, ephremides_age_random, ephremides_age_management, ephremides_age_non_linear, shroff_age_mdp, shroff_age_multi_hop} is that sending updates incurs energy expenditure where energy becomes available intermittently. \cite{yates_age_eh} considers random service time (time for the update to take effect) and \cite{elif_age_eh} considers zero service time; in our work here, we consider a fixed but non-zero service time.



We consider an energy harvesting two-hop network where a source is sending information updates to a destination through a half-duplex relay, see Fig.~\ref{fig_2hop_sys}. The source and the relay use fixed communication rates. Thus, different from \cite{yates_age_eh, elif_age_eh}, they both incur fixed non-zero amounts of transmission delays to deliver their data. Our setting is offline, and the objective is to minimize the total age of information received by the destination within a given communication session time, subject to energy causality constraints at the source and relay nodes, and data causality constraints at the relay node.

We first solve the single-hop version of this problem where the source communicates directly to the destination, with non-zero update transmission delays, extending the offline results in \cite{elif_age_eh}; we observe that introducing non-zero update transmission delays is equivalent to having minimum inter-update time constraints. We then solve the two-hop problem; we first show that it is not optimal for the source to send a new update before the relay finishes forwarding the previous ones, i.e., the relay's data buffer should not contain more than one update packet waiting for service, otherwise earlier arriving packets become stale. Then, we show that the optimal source transmission times are {\it just in time} for the relay to forward the updates, i.e., it is not optimal to let an update wait in the relay's data buffer after being received; it must be directly forwarded. This contrasts the results in \cite{ruiZhangRelay, gunduz2hop} that study throughput maximization in energy harvesting relay networks. In there, throughput-optimal policies are {\it separable} in the sense that the source transmits the most amount of data to the relay regardless of the relay's energy harvesting profile. In our case, the age-optimal policy is not separable; it treats the source and the relay nodes as one combined node that is communicating to the destination. Hence, our single-hop results serve as a building block to find the solution of the two-hop problem.


\section{System Model and Problem Formulation}

A source node acquires measurement updates from some physical phenomenon and sends them to a destination, through the help of a half-duplex relay, during a communication session of duration $T$ time units. Updates need to be sent as {\it timely} as possible; namely, such that the total {\it age of information} is minimized by time $T$. The age of information metric is defined as
\begin{align}
a(t)\triangleq t-U(t),\quad\forall t
\end{align}
where $U(t)$ is the time stamp of the latest received update packet at the destination, i.e., the time at which it was acquired at the source. Without loss of generality, we assume $a(0)=0$. The objective is to minimize the following quantity
\begin{align}
A_T\triangleq\int_{0}^Ta(t)dt
\end{align}

Both the source and the relay depend on energy harvested from nature to transmit their data, and are equipped with infinite-sized batteries to save their incoming energy. Energy arrives in packets of amounts $E$ and $\bar{E}$ at the source and the relay, respectively. Update packets are of equal length, and are transmitted using {\it fixed} rates at the source and the relay. We assume that one update transmission consumes one energy packet at a given node, and hence the number of updates is equal to the minimum of the number of energy arrivals at the source and the relay. Under a fixed rate policy, each update takes $d$ and $\bar{d}$ amount of time to get through the source-relay channel and the relay-destination channel, respectively\footnote{$d$ can be considered, for instance, equal to $B/r$ where $B$ is the update packet length in bits and $r=g(E)$ is the transmission rate in bits/time units, where $g$ is some increasing function representing the rate-energy relationship.}.

Source energy packets arrive at times $\{s_1,s_2,\dots,s_N\}\triangleq{\bm s}$, and relay energy packets arrive at times $\{\bar{s}_1,\bar{s}_2,\dots,\bar{s}_N\}\triangleq\bar{\bm s}$, where without loss of generality we assume that both the source and the relay receive $N$ energy packets, since each update consumes one energy packet in transmission from either node, and hence any extra energy arrivals at either the source or the relay cannot be used. Let $t_i$ and $\bar{t}_i$ denote the transmission time of the $i$th update at the source and the relay, respectively. We first impose the following constraints
\begin{align}
t_i\geq s_i,~\bar{t}_i\geq\bar{s}_i,\qquad 1\leq i\leq N \label{eq_en_caus}
\end{align}
representing the {\it energy causality constraints} \cite{jingP2P} at the source and the relay, which mean that no energy packet can be used before being harvested. Next, we must have
\begin{align}
t_i+d\leq\bar{t}_i,\quad 1\leq i\leq N \label{eq_data_caus}
\end{align}
to ensure that the relay does not forward an update before receiving it from the source, which represents the {\it data causality constraints} \cite{jingP2P}. We also have the {\it service time constraints}
\begin{align}
t_i+d\leq t_{i+1},~\bar{t}_i+\bar{d}\leq\bar{t}_{i+1},\qquad1\leq i\leq N-1 \label{eq_1_tx}
\end{align}
which ensure that there can only be one transmission at a time at the source and the relay. Hence, $d$ and $\bar{d}$ represent the service (busy) time of the source and relay servers, respectively.

\begin{figure}[t]
\center
\includegraphics[scale=.7]{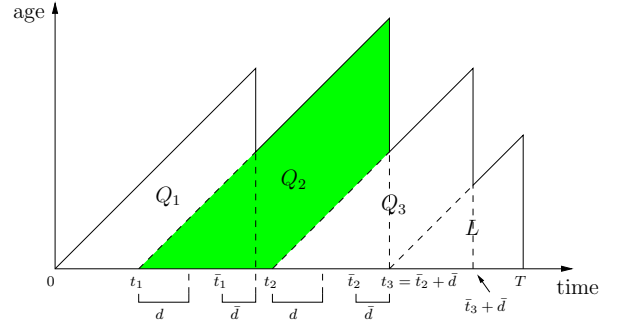}
\caption{Age evolution in a two-hop network with three updates.}
\label{fig_age_2hop}
\vspace{-0.2in}
\end{figure}

Transmission times at the source and the relay should also be related according to the half-duplex nature of the relay operation. For that, we must have the {\it half-duplex constraints}
\begin{align}
(t_i,t_i+d)\cap(\bar{t}_j,\bar{t}_j+\bar{d})=\emptyset,\quad \forall i,j \label{eq_hf_dp_orig}
\end{align}
where $\emptyset$ denotes the empty set, since the relay cannot receive and transmit simultaneously. These constraints enforce that either the source transmits a new update after the relay finishes forwarding the prior one, i.e., $t_{i+1}\geq\bar{t}_i+\bar{d}$ for some $i$; or that the source delivers a new update before the relay starts transmitting the prior one, i.e., $t_{i+k}+d\leq\bar{t}_i$ for some $i$ and $k$. The latter case means that there are $k+1$ update packets waiting in the relay's data buffer just before time $\bar{t}_i$. We prove that this case is not age-optimal. To see this, consider the example of having $k+1=2$ update packets in the relay's data buffer waiting for service. The relay in this case has two choices at its upcoming transmission time: 1) forward the first update followed by the second one sometime later, or 2) forward the second update only and ignore the first one. These two choices yield different age evolution curves. We observe, geometrically, that $A_T$ under choice 2 is strictly less than that under choice 1. Since the source under choice 2 consumes an extra energy packet to send the first update unnecessarily, it should instead save this energy packet to send a new update after the first one is forwarded by the relay. Therefore, it is optimal to replace the half-duplex constraints in (\ref{eq_hf_dp_orig}) by the following reduced ones
\begin{align}
\bar{t}_i+\bar{d}\leq t_{i+1},\quad 1\leq i\leq N-1 \label{eq_hf_dp}
\end{align}

Next, observe that (\ref{eq_1_tx}) can be removed from the constraints since it is implied by (\ref{eq_data_caus}) and (\ref{eq_hf_dp}). In conclusion, the constraints are now those in (\ref{eq_en_caus}), (\ref{eq_data_caus}), and (\ref{eq_hf_dp}). 

Finally, we add the following constraint to ensure reception of all updates by time $T$
\begin{align}
\bar{t}_N+\bar{d}\leq T
\end{align}

In Fig.~\ref{fig_age_2hop}, we present an example of the age of information in a system with 3 updates. The area under the curve representing $A_T$ is given by the sum of the areas of the trapezoids $Q_1$, $Q_2$, and $Q_3$, in addition to the area of the triangle $L$. The area of $Q_2$ for instance is given by $\frac{1}{2}\left(\bar{t}_2+\bar{d}-t_1\right)^2-\frac{1}{2}\left(\bar{t}_2+\bar{d}-t_2\right)^2$.
The objective is to choose feasible transmission times for the source and the relay such that $A_T$ is minimized. Computing the area under the age curve for general $N$ arrivals, we formulate the problem as follows
\begin{align} \label{opt_main}
\min_{{\bm t},\bar{\bm t}}\quad&\sum_{i=1}^N\left(\bar{t}_i+\bar{d}-t_{i-1}\right)^2-\left(\bar{t}_i+\bar{d}-t_i\right)^2 + \left(T-t_N\right)^2 \nonumber \\
\mbox{s.t.}\quad&t_i\geq s_i,~\bar{t}_i\geq\bar{s}_i,\quad1\leq i\leq N \nonumber \\
&t_i+d\leq\bar{t}_i,\quad1\leq i\leq N\nonumber\\
&\bar{t}_i+\bar{d}\leq t_{i+1},\quad1\leq i\leq N
\end{align}
with $t_0\triangleq0$ and $t_{N+1}\triangleq T$.

We note that the energy arrival times ${\bm s}$ and $\bar{\bm s}$, the transmission delays $d$ and $\bar{d}$, the session time $T$, and the number of energy arrivals $N$, are such that problem (\ref{opt_main}) has a feasible solution. This is true only if
\begin{align}
T&\geq\bar{s}_i+\left(N-i+1\right)\bar{d},\quad\forall i \label{eq_feas_rl}\\
T&\geq s_i+\left(N-i+1\right)\left(d+\bar{d}\right),\quad\forall i \label{eq_feas_s}
\end{align}
where (\ref{eq_feas_rl}) (resp. (\ref{eq_feas_s})) ensures that the $i$th energy arrival time at the relay (resp. source) is small enough to allow the reception of the upcoming $N-i$ updates within time $T$. 

\section{Solution Building Block:\\The Single-User Channel}

\begin{figure}[t]
\center
\includegraphics[scale=.8]{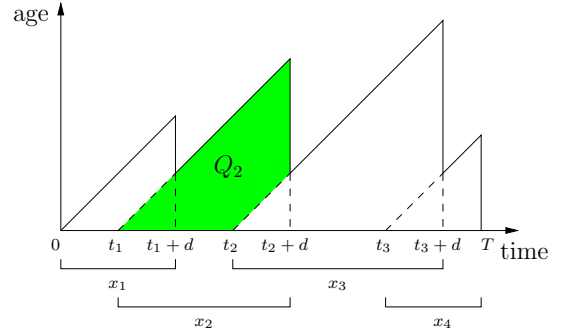}
\caption{Age evolution using in a single-user channel with three updates.}
\label{age_su}
\vspace{-0.2in}
\end{figure}

In this section, we solve the single-user version of problem (\ref{opt_main}); namely, when the source is communicating directly with the destination. We use the solution to the single-user problem in this section as a building block to solve problem (\ref{opt_main}) in the next section. In Fig.~\ref{age_su}, we show an example of the age evolution in a single-user setting. The area of $Q_2$ is now given by $\frac{1}{2}\left(t_2+d-t_1\right)^2-\frac{1}{2}d^2$. We compute the area under the age curve for general $N$ arrivals and formulate the single-user problem as follows
\begin{align} \label{opt_su}
\min_{{\bm t}}\quad&\sum_{i=1}^N\left(t_i+d-t_{i-1}\right)^2+\left(T-t_N\right)^2\nonumber \\
\mbox{s.t.}\quad&t_i\geq s_i,\quad 1\leq i\leq N \nonumber \\
&t_i+d\leq t_{i+1},\quad 1\leq i\leq N
\end{align}
where the second constraints are the service time constraints.

We note that reference \cite{elif_age_eh} considered problem (\ref{opt_su}) when the transmission delay $d=0$. We extend their results for a positive delay (and hence a finite transmission rate) in this section. We first introduce the following change of variables: $x_1\triangleq t_1+d$; $x_i\triangleq t_i-t_{i-1}+d,~2\leq i\leq N$; and $x_{N+1}\triangleq T-t_N$. These variables must satisfy $\sum_{i=1}^{N+1}x_i=T+Nd$, which reflects the dependent relationship between the new variables $\{x_i\}$. This can also be seen from Fig.~\ref{age_su}. Substituting by $\{x_i\}$ in problem (\ref{opt_su}), we get the following equivalent problem
\begin{align} \label{opt_su_eq}
\min_{{\bm x}}\quad&\sum_{i=1}^{N+1}x_i^2 \nonumber\\
\mbox{s.t.}\quad&\sum_{i=1}^kx_i\geq s_k+kd,\quad1\leq k\leq N \nonumber\\
&x_i\geq2d,\quad2\leq i\leq N \nonumber\\
&x_{N+1}\geq d \nonumber\\
&\sum_{i=1}^{N+1}x_i=T+Nd
\end{align}
Observe that problem (\ref{opt_su_eq}) is a convex problem that can be solved by standard techniques \cite{boyd}. For instance, we introduce the following Lagrangian
\begin{align}
\mathcal{L}=&\sum_{i=1}^{N+1}x_i^2-\sum_{k=1}^N\lambda_k\left(\sum_{i=1}^kx_i- s_k-kd\right)-\sum_{i=2}^N\eta_i\left(x_i-2d\right)\nonumber\\
&-\eta_{N+1}\left(x_{N+1}-d\right)+\nu\left(\sum_{i=1}^{N+1}x_i-T-Nd\right)
\end{align}
where $\{\lambda_1,\dots,\lambda_N,\eta_2\dots,\eta_{N+1},\nu\}$ are Lagrange multipliers, with $\lambda_i,\eta_i\geq0$ and $\nu\in\mathbb{R}$. Differentiating with respect to $x_i$ and equating to 0 we get the following KKT conditions
\begin{align}
x_1&=\sum_{k=1}^N\lambda_k-\nu\label{eq_kkt_1}\\
x_i&=\sum_{k=i}^N\lambda_k+\eta_i-\nu,\quad2\leq i\leq N\label{eq_kkt_2}\\
x_{N+1}&=\eta_{N+1}-\nu\label{eq_kkt_3}
\end{align}
along with complementary slackness conditions
\begin{align}
\lambda_k\left(\sum_{i=1}^kx_i- s_k-kd\right)&=0,\quad1\leq k\leq N\label{eq_slk_1} \\
\eta_i\left(x_i-2d\right)&=0,\quad1\leq i\leq N\label{eq_slk_2} \\
\eta_{N+1}(x_{N+1}-d)&=0\label{eq_slk_3}
\end{align}

We now have the following lemmas characterizing the optimal solution of problem (\ref{opt_su_eq}): $\{x_i^*\}$. Lemmas~\ref{thm_x_i_dec} and \ref{thm_x_N_N1} show that the sequence $\{x_i^*\}_{i=2}^{N+1}$ is non-increasing, and derive necessary conditions for it to strictly decrease. On the other hand, Lemma~\ref{thm_x_12} shows that $x_1^*$ can be smaller or larger than $x_2^*$, and derives necessary conditions for the two cases.

\begin{lemma} \label{thm_x_i_dec}
For $2\leq i\leq N-1$, $x_i^*\geq x_{i+1}^*$. Furthermore, $x_i^*> x_{i+1}^*$ only if $\sum_{j=1}^ix_j^*=s_i+id$.
\end{lemma}

\begin{Proof}
We show this by contradiction. Assume that for some $i\in\{2,\dots,N-1\}$ we have $x_i^*<x_{i+1}^*$. By (\ref{eq_kkt_2}), this is equivalent to having $\lambda_i+\eta_i<\eta_{i+1}$, i.e., $\eta_{i+1}>0$, which implies by complementary slackness in (\ref{eq_slk_2}) that $x_{i+1}^*=2d$. This means that $x_i^*<2d$, i.e., infeasible. Therefore $x_i^*\geq x_{i+1}^*$ holds. This proves the first part of the lemma.

To show the second part, observe that since $x_i^*>x_{i+1}^*$ holds if and only if $\lambda_i+\eta_i>\eta_{i+1}$, then either $\lambda_i>0$ or $\eta_i>0$. If $\eta_i>0$, then by (\ref{eq_slk_2}) we must have $x_i^*=2d$, which renders $x_{i+1}^*<2d$, i.e., infeasible. Therefore, $\eta_i$ cannot be positive and we must have $\lambda_i>0$. By complementary slackness in (\ref{eq_slk_1}), this implies that $\sum_{j=1}^ix_j^*=s_i+id$.
\end{Proof}

\begin{lemma} \label{thm_x_12}
$x_1^*>x_2^*$ only if $x_1^*=s_1+d$; while $x_1^*<x_2^*$ only if $x_i^*=2d$, for $2\leq i\leq N$.
\end{lemma}

\begin{Proof}
The necessary condition for $x_1^*$ to be larger than $x_2^*$ can be shown using the same arguments as in the proof of the second part of Lemma~\ref{thm_x_i_dec}, and is omitted for brevity. Let us now assume that $x_1^*$ is smaller than $x_2^*$. By (\ref{eq_kkt_1}) and (\ref{eq_kkt_2}), this occurs if and only if $\eta_2>\lambda_1$, which implies that $x_2^*=2d$ by complementary slackness in (\ref{eq_slk_2}). Finally, by Lemma~\ref{thm_x_i_dec}, we know that $\{x_i^*\}_{i=2}^N$ is non-increasing; since they are all bounded below by $2d$, and $x_2^*=2d$, then they must all be equal to $2d$.
\end{Proof}

\begin{lemma} \label{thm_x_N_N1}
$x_N^*\geq x_{N+1}^*$. Furthermore, $x_N^*>x_{N+1}^*$ only if at least: 1) $\sum_{i=1}^Nx_i^*=s_N+Nd$, or 2) $x_N^*=2d$ occurs.
\end{lemma}

The proof of Lemma~\ref{thm_x_N_N1} is along the same lines of the proofs of the previous two lemmas and is omitted for brevity.

We will use the results of Lemmas~\ref{thm_x_i_dec}, \ref{thm_x_12}, and \ref{thm_x_N_N1} to derive the optimal solution of problem (\ref{opt_su_eq}). To do so, one has to consider the relationship between the parameters of the problem: $T$, $d$, and $N$. For instance, one expects that if the session time $T$ is much larger than the minimum inter-update time $d$, then the energy causality constraints will be binding while the constraints enforcing one update at a time will not be, and vice versa. We formalize this idea by considering two different cases as follows.

\subsection{$Nd\leq T<(N+1)d$}

We first note that $Nd$ is the least value that $T$ can have for problem (\ref{opt_su_eq}) to admit a feasible solution. In this case, the following theorem shows that the optimal solution is achieved by sending all updates back to back with the minimal inter-update time possible to allow the reception of all of them by the end of the relatively small session time $T$.

\begin{theorem} \label{thm_T_sml}
Let $Nd\leq T<(N+1)d$. Then, the optimal solution of problem (\ref{opt_su_eq}) is given by
\begin{align}
x_1^*&=\max\left\{\frac{T-(N-2)d}{2},\max_{1\leq k\leq N}\left\{s_k-\left(k-2\right)d\right\}\right\} \label{eq_x1_T_sml}\\
x_i^*&=2d, \quad 2\leq i\leq N \label{eq_xi_T_sml}\\
x_{N+1}^*&=T-(N-2)d-x_1^* \label{eq_x_N1_T_sml}
\end{align}
\end{theorem}

\begin{Proof}
We first argue that if $x_1^*\geq x_2^*(\geq2d)$, then $\sum_{i=1}^{N+1}x_i^*\geq(2N+1)d$. The last constraint in problem (\ref{opt_su_eq}) then implies that $T\geq(N+1)d$, which is infeasible in this case. Therefore, we must have $x_1^*<x_2^*$. By Lemma~\ref{thm_x_12}, this occurs only if $x_i^*=2d$ for $2\leq i\leq N$. Hence, we set $x_{N+1}=T-(N-2)d-x_1$, and observe that problem (\ref{opt_su_eq}) in this case reduces to a problem in only one variable $x_1$ as follows
\begin{align}
\min_{x_1}\quad &x_1^2+\left(T-(N-2)d-x_1\right)^2 \nonumber \\
\mbox{s.t.}\quad &\max_{1\leq k\leq N}\left\{s_k-(k-2)d\right\}\leq x_1\leq T-(N-1)d
\end{align}
whose solution is given by projecting the critical point of the objective function onto the feasible interval since the problem is convex \cite{boyd}. This directly gives (\ref{eq_x1_T_sml}).
\end{Proof}

\subsection{$T\geq (N+1)d$} \label{sec_su_largeT}

In this case, we propose an algorithmic solution that is based on the necessary optimality conditions in Lemmas~\ref{thm_x_i_dec}, \ref{thm_x_12}, and \ref{thm_x_N_N1}. We first solve problem (\ref{opt_su_eq}) without considering the service time constraints, i.e., assuming that the set of constraints $\{x_i\geq 2d,~2\leq i\leq N;~x_{N+1}\geq d\}$ is not active. We then check if any of these abandoned constraints is not satisfied, and optimally alter the solution to make it feasible.

Let us denote by $\left(P^e\right)$ problem (\ref{opt_su_eq}) without the set of constraints $\{x_i\geq 2d,~2\leq i\leq N;~x_{N+1}\geq d\}$, i.e., considering only the energy causality constraints. We then introduce the following algorithm to solve problem $\left(P^e\right)$

\begin{definition}[Inter-Update Balancing Algorithm] \label{def_alg_eq}
Start by computing
\begin{align}
i_1\triangleq\arg\max\left\{s_1,\frac{s_2}{2},\dots,\frac{s_N}{N},\frac{T-d}{N+1}\right\}
\end{align}
where the set is indexed as $\{1,\dots,N+1\}$, and then set
\begin{align}
x_1^*=\dots=x_{i_1}^*=\max\left\{s_1,\frac{s_2}{2},\dots,\frac{s_N}{N},\frac{T-d}{N+1}\right\}+d
\end{align}
If $i_1=N+1$ stop, else compute

\begin{align}
i_2\triangleq\arg\max\left\{s_{i_1+1}-s_{i_1},\frac{s_{i_1+2}-s_{i_1}}{2},\dots,\right.\nonumber \\
\left.\frac{s_N-s_{i_1}}{N-i_1},\frac{T-d-s_{i_1}}{N+1-i_1}\right\}
\end{align}
where the set is indexed as $\{i_1+1,\dots,N+1\}$, and then set
\begin{align}
x_{i_1+1}^*=\dots=x_{i_2}^*=&\max\left\{s_{i_1+1}-s_{i_1},\frac{s_{i_1+2}-s_{i_1}}{2},\dots,\right.\nonumber \\
&\left.\frac{s_N-s_{i_1}}{N-i_1},\frac{T-d-s_{i_1}}{N+1-i_1}\right\}+d
\end{align}
If $i_2=N+1$ stop, else continue with computing $i_3$ as above. The algorithm is guaranteed to stop since it will at most compute $i_{N+1}$ which is equal to $N+1$ by construction.
\end{definition}

Note that while computing $i_k$, if the $\arg\max$ is not unique, we pick the largest maximizer. Observe that the algorithm equalizes the $x_i$'s as much as allowed by the energy causality constraints. Let $\{\bar{x}_i\}_{i=1}^N$ be the output of the Inter-Update Balancing algorithm and let $\{x_i^e\}_{i=1}^N$ denote the optimal solution of problem $(P^e)$. We now have the following results

\begin{lemma} \label{thm_eq_alg_prop}
$\{\bar{x}_i\}_{i=1}^N$ is a non-increasing sequence, and $\bar{x}_j>\bar{x}_{j+1}$ only if $\sum_{i=1}^j\bar{x}_i=s_j+jd$.
\end{lemma}

\begin{lemma} \label{thm_eq_alg_opt}
$x_i^e=\bar{x}_i$, $1\leq i\leq N$.
\end{lemma}

Lemma~\ref{thm_eq_alg_prop} can be shown using contradiction arguments and the definition of $i_k$. Lemma~\ref{thm_eq_alg_opt} is similar to \cite[Theorem 1]{elif_age_eh}. In fact, the Inter-Update Balancing algorithm reduces to the optimal offline algorithm proposed in \cite{elif_age_eh} when $d=0$. When $d>0$, some change of parameters can still show the equivalence. The complete proofs of the two lemmas are omitted due to space limits. The next corollary now follows.

\begin{corollary} \label{thm_partial_opt}
Consider problem $(P^e)$ with the additional constraint that $\sum_{i=1}^jx_i=s_j+jd$ holds for some $j\leq N$. Then, the optimal solution of the problem, under this condition, for time indices not larger than $j$ is given by $\{x_i^e\}_{i=1}^j$.
\end{corollary}

 
The following theorem shows that the optimal solution of problem (\ref{opt_su_eq}), $\{x_i^*\}$, is found by equalizing the inter-update times as much as allowed by the energy causality constraints. If such equalization does not satisfy the minimal inter-update time constraints, we force it to be exactly equal to such minimum and adjust the last variable $x_{N+1}$ accordingly.

\begin{theorem} \label{thm_amnd}
Let $T\geq(N+1)d$. If $x_i^e\geq2d,~2\leq i\leq N$ and $x_{N+1}^e\geq d$, then $x_i^*=x_i^e,~\forall i$. Else, let $n_0$ be the first time index at which $\{x_i^e\}$ is not feasible in problem (\ref{opt_su_eq}). Then, we have $n_0\leq N$. If $n_0>2$, we have
\begin{align}
x_i^*&=x_i^e,\quad1\leq i\leq n_0-1 \\
x_i^*&=2d,\quad n_0\leq i\leq N \\
x_{N+1}^*&=T+Nd-\sum_{i=1}^Nx_i^*
\end{align}
Otherwise, for $n_0=2$, $\{x_i^*\}$ is given by the above if $x_1^e=s_1+d$, else $\{x_i^*\}$ is given by (\ref{eq_x1_T_sml})-(\ref{eq_x_N1_T_sml}).
\end{theorem}

\begin{Proof}
The first part of the theorem follows directly since the solution of the less constrained problem $\left(P^e\right)$ is optimal if feasible in problem (\ref{opt_su_eq}). Next, we prove the second part.

We first show that $n_0\leq N$ by contradiction. Assume that $n_0=N+1$, i.e., $x_{N+1}^e<d$ and $x_N^e\geq 2d>x_{N+1}^e$. By Lemma~\ref{thm_eq_alg_prop}, this means that $\sum_{i=1}^Nx_i^e=s_N+Nd$. Hence, $x_{N+1}^e=T+Nd-s_N-Nd=T-s_N$, which cannot be less than $d$ by the feasibility assumption in (\ref{eq_feas_rl}). Thus, $n_0\leq N$.

Now let $n_0>2$ and observe that $x_{n_0}^e<2d\leq x_{n_0-1}$. Thus, by Lemma~\ref{thm_eq_alg_prop}, we must have $\sum_{i=1}^{n_0-1}x_i^e=s_{n_0-1}+(n_0-1)d$. Now let us show that the proposed policy is feasible; we only need to check whether $x_{N+1}^*\geq d$. Towards that, we have
\begin{align}
x_{N+1}^*&=T+Nd-\sum_{i=1}^{n_0-1}x_i^*-\left(N-n_0+1\right)2d \nonumber \\
&=T-s_{n_0-1}-(N-n_0+1)d \geq d
\end{align}
where the last inequality follows by the feasibility assumption in (\ref{eq_feas_rl}). Therefore, the proposed policy is feasible.

We now show that it is optimal as follows. Assume that there exists another policy $\{\tilde{x}_i\}$ that achieves a lower age than $\{x_i^*\}$. We now have two cases. First, assume that $\sum_{i=1}^{n_0-1}\tilde{x}_i=s_{n_0-1}+(n_0-1)d$. then by Corollary~\ref{thm_partial_opt} we must have $\tilde{x}_i=x_i^*$ for $1\leq i\leq n_0-1$. Now for $n_0\leq i\leq N$, if $\tilde{x}_i>x_i^*=2d$, this means that $\tilde{x}_{N+1}<x_{N+1}^*$ to satisfy the last constraint in (\ref{opt_su_eq}). Since $\sum_{i=n_0}^{N+1}\tilde{x}_i=\sum_{i=n_0}^{N+1}x_i^*$, then by convexity of the square function, $\sum_{i=n_0}^{N+1}\left(\tilde{x}_i\right)^2>\sum_{i=n_0}^{N+1}\left(x_i^*\right)^2$ \cite{boyd}, and hence $\{\tilde{x}_i\}$ cannot be optimal. Second, assume that $\sum_{i=1}^{n_0-1}\tilde{x}_i>s_{n_0-1}+(n_0-1)d=\sum_{i=1}^{n_0-1}x_i^*$. Since $\tilde{x}_i\geq x_i^*=2d$ for $n_0\leq i\leq  N$, and $\sum_{i=1}^{N+1}\tilde{x}_i=\sum_{i=1}^{N+1}x_i^*$, then we must have $\tilde{x}_{N+1}<x_{N+1}^*$. Thus, $\sum_{i=1}^{N+1}\left(\tilde{x}_i\right)^2>\sum_{i=1}^{N+1}\left(x_i^*\right)^2$ by convexity of the square function \cite{boyd}, and $\{\tilde{x}_i\}$ cannot be optimal.

Finally, let $n_0=2$. If $x_1^e=s_1+d$, then the proof follows by the arguments for the $n_0>2$ case. Else if $x_1^e>s_1+d$, then $x_1^e=x_2^e\geq x_{N+1}^e$ by Lemma~\ref{thm_eq_alg_prop}. Since $\{x_i^e\}_{i=2}^N$ have to increase to at least $2d$, then $x_1^e+x_{N+1}^e$ has to decrease to satisfy the last constraint in (\ref{opt_su_eq}). However, one cannot increase $x_1^e$ to $2d$ or more and compensate that by decreasing $x_{N+1}^e$, by convexity of the square function. Thus, $x_1^*<x_2^*$, and Lemma~\ref{thm_x_12} shows that the results of Theorem~\ref{thm_T_sml} follow to give (\ref{eq_x1_T_sml})-(\ref{eq_x_N1_T_sml}).
\end{Proof}

\section{Two-Hop Network:\\Solution of Problem (\ref{opt_main})}

We now discuss how to use the results of the single-user problem to solve problem (\ref{opt_main}). We have the following theorem.

\begin{theorem} \label{thm_2hop}
The optimal solution of problem (\ref{opt_main}) is given by the optimal solution of problem (\ref{opt_su}) after replacing $s_i$ by $\max\{\bar{s}_i,s_i+d\},~\forall i$; $d$ by $d+\bar{d}$; and $T$ by $T+d$.
\end{theorem}
\vspace{-0.1in}
\begin{Proof}
Let $f$ denote the objective function of problem (\ref{opt_main}). Differentiating $f$ with respect to $t_i$, $i\leq N-1$, we get $\frac{\partial f}{\partial t_i}=2\left(\bar{t}_i+\bar{d}-t_i\right)-2\left(\bar{t}_{i+1}+\bar{d}-t_i\right)$,
which is negative since $\bar{t}_{i+1}>\bar{t}_i$. We also have $\frac{\partial f}{\partial t_N}=2\left(\bar{t}_N+\bar{d}-t_N\right)-2\left(T-t_N\right)$, which is non-positive since $\bar{t}_N+\bar{d}\leq T$. Thus, $f$ is decreasing in $\{t_i\}_{i=1}^{N-1}$ and non-increasing in $t_N$. Therefore, the optimal $\{t_i^*\}$ satisfies the data causality constraints in (\ref{eq_data_caus}) with equality for all updates so as to be the largest possible and achieve the smallest $A_T$. Setting $t_i=\bar{t}_i-d,~\forall i$ in problem (\ref{opt_main}) we get
\vspace{-0.05in}
\begin{align}
f=\sum_{i=1}^N\left(\bar{t}_i+\bar{d}+d-\bar{t}_{i-1}\right)^2-N\left(\bar{d}+d\right)^2 + \left(T+d-\bar{t}_N\right)^2 
\end{align}
with the constraints now being
\begin{align}
&\bar{t}_i\geq s_i+d,~\bar{t}_i\geq\bar{s}_i,\quad\forall i \\
&\bar{t}_i+\bar{d}+d\leq\bar{t}_{i+1},\quad 1\leq i\leq N-1 \\
&\bar{t}_N+\bar{d}\leq T
\end{align}
We now see that minimizing $f$ subject to the above constraints is exactly the same as solving problem (\ref{opt_su}) after applying the change of parameters mentioned in the theorem.
\end{Proof}

Theorem~\ref{thm_2hop} shows that the source should send its updates {\it just in time} as the relay is ready to forward, and no update should wait for service in the relay's data buffer. Thus, the source and the relay act as one combined node that can send updates whenever it receives combined energy packets at times $\left\{\max\{\bar{s}_i,s_i+d\}\right\}$. This fundamental observation can be generalized to multi-hop networks as well. Given $M>1$ relays, each node should send updates just in time as the following node is ready to forward, until reaching destination.

\vspace{-0.025in}

\section{Numerical Results}

We now present some numerical examples to further illustrate our results. A two-hop network has energy arriving at times ${\bm s}=[2,6,7,11,13]$ at the source, and $\bar{{\bm s}}=[1,4,9,10,15]$ at the relay. A source transmission takes $d=1$ time unit to reach the relay; a relay transmission takes $\bar{d}=2$ time units to reach the destination. Session time is $T=19$. We apply the change of parameters in Theorem~\ref{thm_2hop} to get new energy arrival times ${\bm s}=[3,7,9,12,15]$, new transmission delay $d=3$, and new session time $T=20$. Then, we solve problem (\ref{opt_su_eq}) to get the optimal inter-update times, using the new parameters. Note that $T\geq(N+1)d=18$, whence the optimal solution is given by Theorem~\ref{thm_amnd}. We apply the Inter-Update Balancing algorithm to get ${\bm x}^e=[6.5,6.5,5.67,5.67,5.67,5]$. Hence, the first infeasible inter-update time occurs at $n_0=3$ ($x_3^e<2d=6$). Thus, we set: $x_1^*=x_1^e$ and $x_2^*=x_2^e$; $x_3^*=x_4^*=x_5^*=2d$; and $x_6^*=T+Nd-\sum_{i=1}^5x_i^*$. We see that ${\bm x}^*=[6.5,6.5,6,6,6,4]$ satisfies the conditions stated in Lemmas~\ref{thm_x_i_dec}, \ref{thm_x_12}, and \ref{thm_x_N_N1}.

We consider another example where energy arrives at times ${\bm s}=[0,4,4,9,13]$ and $\bar{{\bm s}}=[1,3,6,10,12]$, with $T=16$. Applying the change of parameters in Theorem~\ref{thm_2hop} we get $T=17<(N+1)d=18$, and hence we use the results of Theorem~\ref{thm_T_sml} to get ${\bm x}^*=[5,6,6,6,6,3]$. We then increase $T$ to $18$. This is effectively $19$ according to Theorem~\ref{thm_2hop}, and therefore we apply Theorem~\ref{thm_amnd} results. The Inter-Update Balancing algorithm gives ${\bm x}^e=[5.8,5.8,5.8,5.8,5.8,5]$, and hence $n_0=2$. Since $x_1^e>s_1+d=4$, then the optimal solution is given by (\ref{eq_x1_T_sml})-(\ref{eq_x_N1_T_sml}) as ${\bm x}^*=[5,6,6,6,6,5]$.


\section{Conclusions}

We proposed age-minimal policies in energy harvesting two-hop networks with fixed transmission delays. The optimal policy is such that the relay's data buffer should not contain any packets waiting for service; the source should send an update to the relay just in time as the relay is ready to forward. This let us treat the source and relay nodes as one combined node communicating with the destination node, and reduce the two-hop problem to a single hop one. We solved the single hop problem by balancing inter-update times to the extent allowed by energy arrival times and transmission delays.


\begin{thebibliography}{10}

\bibitem{jingP2P}
J.~Yang and S.~Ulukus.
\newblock Optimal packet scheduling in an energy harvesting communication
  system.
\newblock {\em IEEE Trans. Comm.}, 60(1):220--230, January 2012.

\bibitem{kayaEmax}
K.~Tutuncuoglu and A.~Yener.
\newblock Optimum transmission policies for battery limited energy harvesting
  nodes.
\newblock {\em IEEE Trans. Wireless Comm.}, 11(3):1180--1189, March 2012.

\bibitem{omurFade}
O.~Ozel, K.~Tutuncuoglu, J.~Yang, S.~Ulukus, and A.~Yener.
\newblock Transmission with energy harvesting nodes in fading wireless
  channels: Optimal policies.
\newblock {\em IEEE JSAC}, 29(8):1732--1743, September 2011.

\bibitem{ruiZhangEH}
C.~K. Ho and R.~Zhang.
\newblock Optimal energy allocation for wireless communications with energy
  harvesting constraints.
\newblock {\em IEEE Trans. Signal Proc.}, 60(9):4808--4818, September 2012.

\bibitem{jingBC}
J.~Yang, O.~Ozel, and S.~Ulukus.
\newblock Broadcasting with an energy harvesting rechargeable transmitter.
\newblock {\em IEEE Trans. Wireless Comm.}, 11(2):571--583, February 2012.

\bibitem{omurBC}
O.~Ozel, J.~Yang, and S.~Ulukus.
\newblock Optimal broadcast scheduling for an energy harvesting rechargebale
  transmitter with a finite capacity battery.
\newblock {\em IEEE Trans. Wireless Comm.}, 11(6):2193--2203, June 2012.

\bibitem{elifBC}
M.~A. Antepli, E.~Uysal-Biyikoglu, and H.~Erkal.
\newblock Optimal packet scheduling on an energy harvesting broadcast link.
\newblock {\em IEEE JSAC}, 29(8):1721--1731, September 2011.

\bibitem{jingMAC}
J.~Yang and S.~Ulukus.
\newblock Optimal packet scheduling in a multiple access channel with energy
  harvesting transmitters.
\newblock {\em Journal of Comm. and Networks}, 14(2):140--150, April 2012.

\bibitem{kaya-interference}
K.~Tutuncuoglu and A.~Yener.
\newblock Sum-rate optimal power policies for energy harvesting transmitters in
  an interference channel.
\newblock {\em Journal Comm. Networks}, 14(2):151--161, April 2012.

\bibitem{ruiZhangRelay}
C.~Huang, R.~Zhang, and S.~Cui.
\newblock Throughput maximization for the {G}aussian relay channel with energy
  harvesting constraints.
\newblock {\em IEEE JSAC}, 31(8):1469--1479, August 2013.

\bibitem{gunduz2hop}
D.~Gunduz and B.~Devillers.
\newblock Two-hop communication with energy harvesting.
\newblock In {\em IEEE CAMSAP}, December 2011.

\bibitem{berkDiamond-jour}
B.~Gurakan and S.~Ulukus.
\newblock Cooperative diamond channel with energy harvesting nodes.
\newblock {\em IEEE JSAC}, 34(5):1604--1617, May 2016.

\bibitem{varan_twc_jour}
B.~Varan and A.~Yener.
\newblock Delay constrained energy harvesting networks with limited energy and
  data storage.
\newblock {\em IEEE JSAC}, 34(5):1550--1564, May 2016.

\bibitem{arafa_baknina_twc_dec_proc}
A.~Arafa, A.~Baknina, and S.~Ulukus.
\newblock Energy harvesting two-way channels with decoding and processing
  costs.
\newblock {\em IEEE Trans. Green Comm. and Networking}, 1(1):3--16, March 2017.

\bibitem{yates_age_1}
S.~Kaul, R.~Yates, and M.~Gruteser.
\newblock Real-time status: How often should one update?
\newblock In {\em IEEE Infocom}, March 2012.

\bibitem{yates_age_mac}
R.~Yates and S.~Kaul.
\newblock Real-time status updating: Multiple sources.
\newblock In {\em IEEE ISIT}, July 2012.

\bibitem{ephremides_age_random}
C.~Kam, S.~Kompella, and A.~Ephremides.
\newblock Age of information under random updates.
\newblock In {\em IEEE ISIT}, July 2013.

\bibitem{ephremides_age_management}
M.~Costa, M.~Codreanu, and A.~Ephremides.
\newblock On the age of information in status update systems with packet
  management.
\newblock {\em IEEE Trans. Info. Theory}, 62(4):1897--1910, April 2016.

\bibitem{ephremides_age_non_linear}
A.~Kosta, N.~Pappas, A.~Ephremides, and V.~Angelakis.
\newblock Age and value of information: Non-linear age case.
\newblock In {\em IEEE ISIT}, June 2017.

\bibitem{shroff_age_mdp}
Y.~Sun, E.~Uysal-Biyikoglu, R.~Yates, C.~E. Koksal, and N.~B. Shroff.
\newblock Update or wait: How to keep your data fresh.
\newblock In {\em IEEE Infocom}, April 2016.

\bibitem{shroff_age_multi_hop}
A.~M. Bedewy, Y.~Sun, and N.~B. Shroff.
\newblock Age-optimal information updates in multihop networks.
\newblock In {\em IEEE ISIT}, June, 2017.

\bibitem{yates_age_eh}
R.~D. Yates.
\newblock Lazy is timely: Status updates by an energy harvesting source.
\newblock In {\em IEEE ISIT}, June 2015.

\bibitem{elif_age_eh}
B.~T. Bacinoglu, E.~T. Ceran, and E.~Uysal-Biyikoglu.
\newblock Age of information under energy replenishment constraints.
\newblock In {\em UCSD ITA}, February 2015.

\bibitem{boyd}
S.~P. Boyd and L.~Vandenberghe.
\newblock {\em Convex Optimization}.
\newblock 2004.

\end{thebibliography}
\end{document}